\documentclass{iaus}
\usepackage{graphicx}

\title[Stellar Winds in IC1613] 
{Are the Stellar Winds in IC1613 stronger than expected?}

\author[A. Herrero et al.]   
{A. Herrero$^{1,2}$,
 M. Garcia$^{1,2}$,
 K. Uytterhoeven$^{3}$,
 F. Najarro$^{4}$,
 D.J. Lennon$^{5}$,
 S. Simon-D\'{i}az$^{1,2}$,
 N. Castro$^{1,2}$,
 J. Puls$^6$,
 J.S. Vink$^7$,
 M.A. Urbaneja$^8$,
 \and A. de Koter$^{9,10}$}

\affiliation{$^1$Instituto de Astrof\'{i}sica de Canarias, E38200 La Laguna, Spain; 
$^2$Departamento de Astrof\'{i}sica, Universidad de La Laguna, E38205,  La Laguna, Spain; 
$^3$CEA, Saclay, France; $^4$CAB (CSIC-INTA), Torrej\'on de Ardoz, Spain;
$^5$STScI, Baltimore, USA; $^6$ Univ.-Sternwarte Munich, Germany; $^7$ Armagh Observatory, Armagh, Northern Ireland;
$^8$ IfA, Hawaii, USA; $^9$ Astronomical Institut Anton Pannekoek, Amsterdam, The Netherlands;$^{10}$Astronomical Institut, Utrecht, The Netherlands}

\pubyear{2011}
\volume{IAUS272}  
\pagerange{xxx-xxx}
\jname{Active OB Stars: structure, evolution, mass loss and critical limits}
\editors{C. Neiner, G. Wade, G. Meynet, G. Peters}
\begin{document}

\maketitle

\begin{abstract}
In this poster we present the results of our analyses of three early massive stars in IC 1613, whose spectra  have been observed with VIMOS and analyzed with CMFGEN and FASTWIND. One of the targets resulted a possible LBV and the other two are Of stars with unexpectedly strong winds. The Of stars seem to be strongly contaminated by CNO products.
Our preliminary results may represent a challenge for the theory of stellar atmospheres, but they still have to be confirmed by the analysis of more objects and a more  complete coverage of the parameter space.
\end{abstract}

\firstsection 

\section{Why and how we observed stars in IC1613}
IC 1613 is a Local Group dwarf irregular galaxy with a very low metallicity and on-going massive star formation.
Therefore, it is an ideal place to test theories of stellar atmospheres, winds and evolution, and 
compare their predictions with the actual behaviour of massive stars.
For this reason, we have obtained multiwavelength (UBVRI) photometry of IC 1613 with the WFC attached to the INT@ORM (Garcia et al. 2009) and
have secured VIMOS@VLT spectra of IC 1613 stars selected using this photometry. We used the HR-Blue and HR-Orange gratings, yielding R$\approx$2100, $\lambda$= 3900-7000 \AA~ and SNR$\approx$100 after combining 19 blue and 10 red exposures (see \cite{h10}, for more details).

\section{The nature of V39}
The first object we analyzed was variable V39 in IC1613. The nature of V39 has been debated since its discovery. \cite{sandage71}, in his analysis of IC1613 based on unpublished previous work by Baade, points out that it is the only peculiar variable in the field, due to its inverted beta-Lyrae light curve. 
 Baade never considered it a Cepheid, because it is too bright to fit the P-L relation compared to other Cepheids with the same period.
In our observations, V39 displays P-Cygni profiles in the Balmer and Fe\,{\sc ii} lines at $\lambda\lambda$4924, 5018 and 5168 \AA, as well as He\,{\sc i} 5876 \AA~ (the only He\,{\sc i} line seen in the spectrum; its presence is confirmed because it can be independently seen in the blue and red spectrograms).

The V39 spectrum changes from early A in the blue to late G in the red, but without significant spectral variability  (the lack of important 
spectral variability is particularly remarkable in H$_\alpha$). \cite{manteg02} suggested that V39 is actually the casual superposition of a 
Galactic W Vir star and an IC1613 Red Supergiant. However, this is not consistent with our observations, because the spectrum 
has a unique radial velocity, consistent with the IC1613 systemic velocity and the (small) H$_\alpha$ profile variations are not consistent 
with that of an W Vir object. Moreover, a binary system is not consistent with the observed spectroscopic lack of variability in short timescales.

Our analysis of V39, performed by means of CMFGEN, reveals that its stellar parameters and location on the Hertzsprung-Russell Diagram 
are consistent with a low-luminosity 
LBVc or SgB[e] star (see \cite{h10} for details), which raises the question of how to produce such an object at low metallicities. 
We propose V39 to be an LBVc surrounded by a hot thick disk: the LBV would be responsible 
for the P-Cygni profiles observed in the spectrum and the disk would precess and produce the photometric variability and additional absorption in the Balmer lines.
The model is consistent with  the large reddening of V39 ($>$5 times the average IC1613 internal reddening; unfortunately, no Spitzer images are available) but
has to be confirmed or rejected with new observations.

\section{Of stars in IC 1613}

As a second step, we have selected two Of stars from our sample of IC1613 massive stars. 
The analyses have been carried out using the latest FASTWIND version, that includes N\,{\sc iii} dielectronic
recombination (Rivero Gonz\'alez et al., 2011, in prep.). Although results are still preliminary, there are two points of great interest (to be confirmed):
(a) the stars have higher mass-loss rates and wind momenta than expected for the metallicity of IC 1613; (b) the N and He abundances suggest strong CNO contamination at the surface (see Tab.~\ref{tab1}). While the second one can be expected from theory, that predicts increased mixing at lower metallicities,
the first one is not predicted by the theory of radiatively driven winds. If confirmed, it would indicate an unexpected behaviour of stellar winds at low metallicities.
However, before adopting such a conclusion, more objects have to be analyzed and a more complete  coverage of the parameter space (wind clumping, beta parameter, microturbulenceÉ)  is mandatory. 
\begin{table}
  \begin{center}
  \caption{Results of the quantitative analysis of IC1613 Of stars. Mass loss rates are in units of solar masses per year, and terminal velocities 
  (adopted from V$_\infty$-Z relations) are in km s$^{-1}$}
  \label{tab1}
 {\scriptsize
  \begin{tabular}{lccccccccc}\hline 
{\bf Star ID} & {\bf Sp Type} & {\bf Teff} & {\bf log g} & {\bf R/R$_\odot$} & {\bf log(L/L$_\odot$)} & {$\dot{\bf M}$} & {\bf V$_\infty$} & \bf{Y(He)} & \bf{log(N/H)+12} \\ \hline
62024         &  O5 If             & 38500     & 3.50         &  10.5             &  5.34                  & 1.60$\times$10$^{-6}$& 1490 & 0.18   & 8.00 \\
67559         &  O7 I(f)           & 34000    & 3.40         &  12.3             &  5.25                  & 1.23$\times$10$^{-6}$& 1450 & 0.18   & 7.80 \\ \hline
  \end{tabular}
  }
 \end{center}
\vspace{1mm}
\end{table}

\end{document}